\documentclass[symmetry,article,accept,moreauthors,pdftex]{mdpi} 
\usepackage{longtable}

\firstpage{1} 
\pubvolume{1}
\issuenum{1}
\articlenumber{0}
\pubyear{2021}
\copyrightyear{2021}
\externaleditor{Academic Editor: Kazuharu Bamba %MDPI: Please add Academic Editor if avaliable.
} % For journal Automation, please change Academic Editor to "Communicated by"
\datereceived{} 
\dateaccepted{} 
\datepublished{} 
\hreflink{https://doi.org/} % If needed use \linebreak
\Title{The FUor Star V2493 Cyg (HBC 722)---Eleven Years at Maximum~Brightness}

\TitleCitation{The FUor Star V2493 Cyg (HBC 722)---Eleven Years at Maximum Brightness}

\Author{Evgeni Semkov $^{1,}$*\orcidA{}, Sunay Ibryamov $^{2}$\orcidB{} and Stoyanka Peneva $^{1}$}

\AuthorNames{Evgeni Semkov, Sunay Ibryamov and Stoyanka Peneva}

\AuthorCitation{Semkov, E.; Ibryamov, S.; Peneva, S.}
\address{%
$^{1}$ \quad Institute of Astronomy and National Astronomical Observatory, Bulgarian Academy of Sciences, 72, Tsarigradsko Shose Blvd., 1784 Sofia, Bulgaria; speneva@astro.bas.bg\\
$^{2}$ \quad Department of Physics and Astronomy, Faculty of Natural Sciences, University of Shumen, 115, Universitetska Str., 9712 Shumen, Bulgaria; sibryamov@shu.bg}

\corres{Correspondence: esemkov@astro.bas.bg}
\abstract{
At the time of stellar evolution, young stellar objects go through processes of increased activity and instability.
Star formation takes place in several stages during which the star accumulates enough mass to initiate thermonuclear reactions in the nucleus.
A significant percentage of the mass of Sun-like stars accumulates during periods of increased accretion known as FUor outbursts.
Since we know only about two dozen stars of this type, the study of each new object is very important for our knowledge.
In this paper, we present data from photometric monitoring on a FUor object V2493 Cyg discovered in 2010. 
Our data were obtained in the optical region with BVRI Johnson--Cousins set of filters during the period from November 2016 to February 2021.
The results of our observations show that during this period no significant changes in the brightness of the star were registered.
We only detect variations with a small amplitude around the maximum brightness value.
Thus, since 2013 V2493 Cyg remains at its maximum brightness, without a decrease in brightness. 
Such photometric behavior is not typical of other stars from FUor type.
Usually, the light curves of FUors are asymmetrical, with a very rapid rise and gradual decline of the brightness.
V2493 Cyg remains unique in this respect with a very rapid rise in brightness and prolonged retention in maximum light.
Our period analysis made for the interval February 2013--February 2021 reveals a well-defined period of 914~$\pm$~10 days. 
Such periodicity can be explained by dust structures remaining from star formation in orbit around the star.}


\begin{document}
%%%%%%%%%%%%%%%%%%%%%%%%%%%%%%%%%%%%%%%%%%

\section{Introduction}

The main characteristic of the young stellar objects is their photometric and spectral variability.
In fact, most of the pre-main sequence (PMS) stars show variations in brightness that are associated with the evolutionary processes. 
Most often the variability is observed as transient increases in brightness, outbursts, temporary drops in brightness (eclipses), irregular or
regular variations for a short or long time scales.
Stars that undergo episodic outbursts with large amplitudes can be divided into two types FUors and EXors~\cite{Her89}.
These two types of eruptive stars seem to be related to the low-mass PMS objects (T Tauri stars), which have massive circumstellar disks~\cite{Her77,Har96,Rei10}. 

One of the most remarkable phenomenon in the early stages of stellar evolution is the FUor outbursts.
The prototype of FUors is the eruptive star FU Orionis, located in the Orion star forming region. 
The star was brightened by 6 magnitudes in 1936 and for a long time was the only one object of its kind~\cite{Wac54}. 
First Ambartsumian~\cite{Amb71} draws attention to this object by linking it to the evolution of the young stellar objects and proposes the abbreviation FUor.
FUors were defined as a class of young variables by Herbig~\cite{Her77} after the discovery of two new FUor objects: V1057 Cyg and V1515 Cyg. 
About twenty new FUor objects were assigned to this class of young eruptive stars over the next four decades~\cite{Rei10, Aud14, Con18}.
Due to the very small number of known FUor objects, each newly discovered attracts significant attention.

All known FUors share the same defining characteristics: location in star-forming regions, outburst with an amplitude of about 4--6 magnitudes, association with reflection nebulae, an F-G supergiant spectrum during the outburst, a strong Li I~6707~\AA\ line in absorption, and CO bands in near-infrared spectra~\cite{Her77, Rei10}.
During the outbursts, FUor objects undergo significant increase in their accretion rate from $\sim$10$^{-7}$$M_{\odot}$$/$yr up to $\sim$10$^{-4}$$M_{\odot}$$/$yr.
A typical outburst of FUor objects can last for several decades or a century, and the rise time is usually shorter than that of the decline.

The triggering mechanism of this enhanced accretion rate is carried out by changes in the structure and mass of the circumstellar disk or in the stellar environment.
The most popular is that the outburst is caused by gravitational or thermal instability in the circumstellar disk~\cite{Har96, Zhu09}.
Another possible triggering mechanisms could be the interactions of the circumstellar disk with a planet or nearby stellar companion on an eccentric \linebreak orbit~\cite{Lod04, Rei04, Pfa08} and in fall of clumps of material formed by disk fragmentation onto the central star~\cite{Vor05,Vor21}.   
For a period of $\sim$100 years, the circumstellar disk adds $\sim$$10^{-2}$ $M_{\odot}$ onto the central star and ejects $\sim$10\% of the accreting material in a high-velocity stellar wind. 
During the early evolutionary stages of the solar mass stars they probably went through several dozen episodes of such increased accretion. 
Calculations show that up to 50$\%$ of the protostellar mass can be accumulated due to the FUor phenomenon.
 
\begin{figure}[H]
\includegraphics[width=6.6 cm]{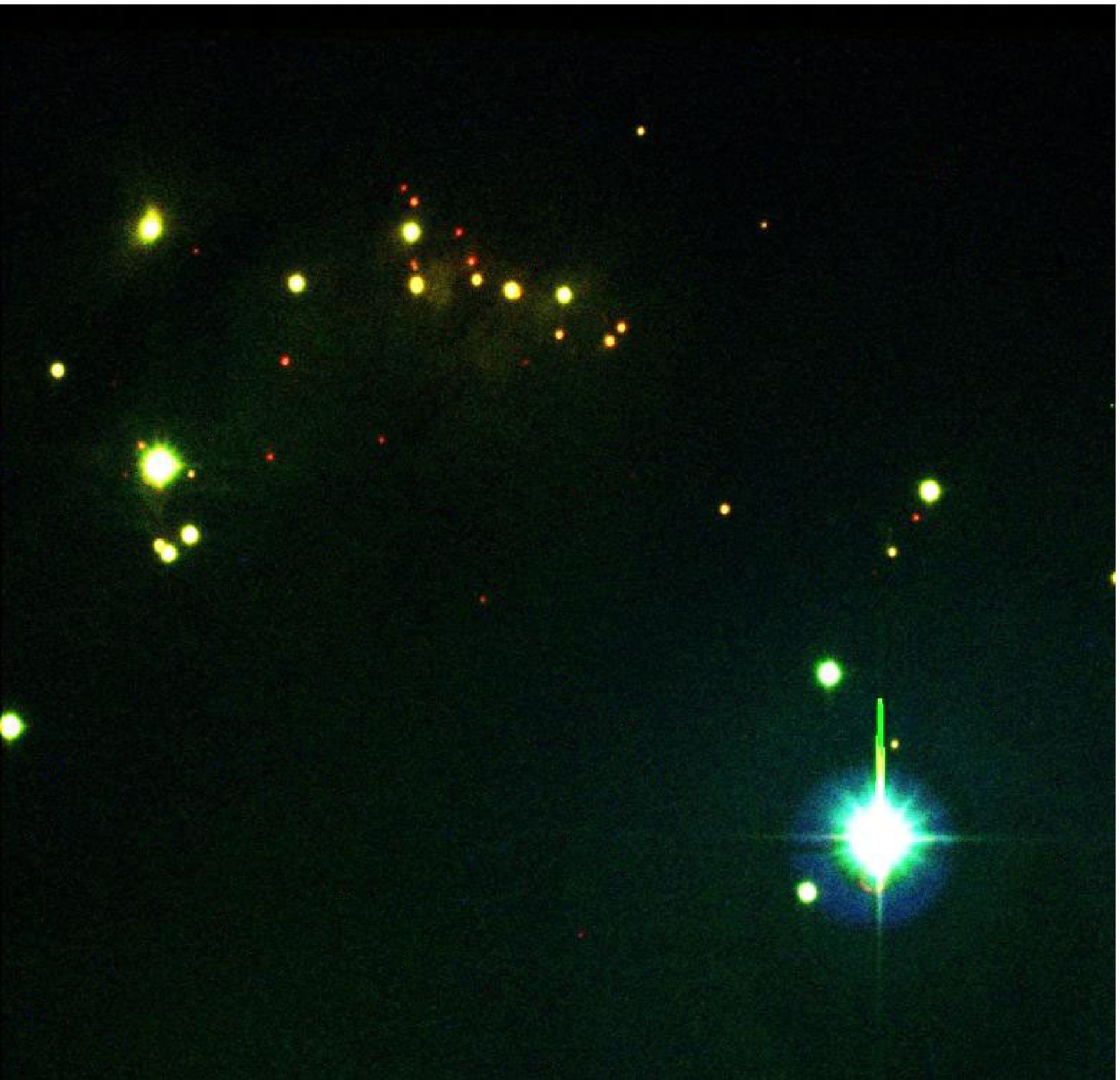}
\includegraphics[width=6.6 cm]{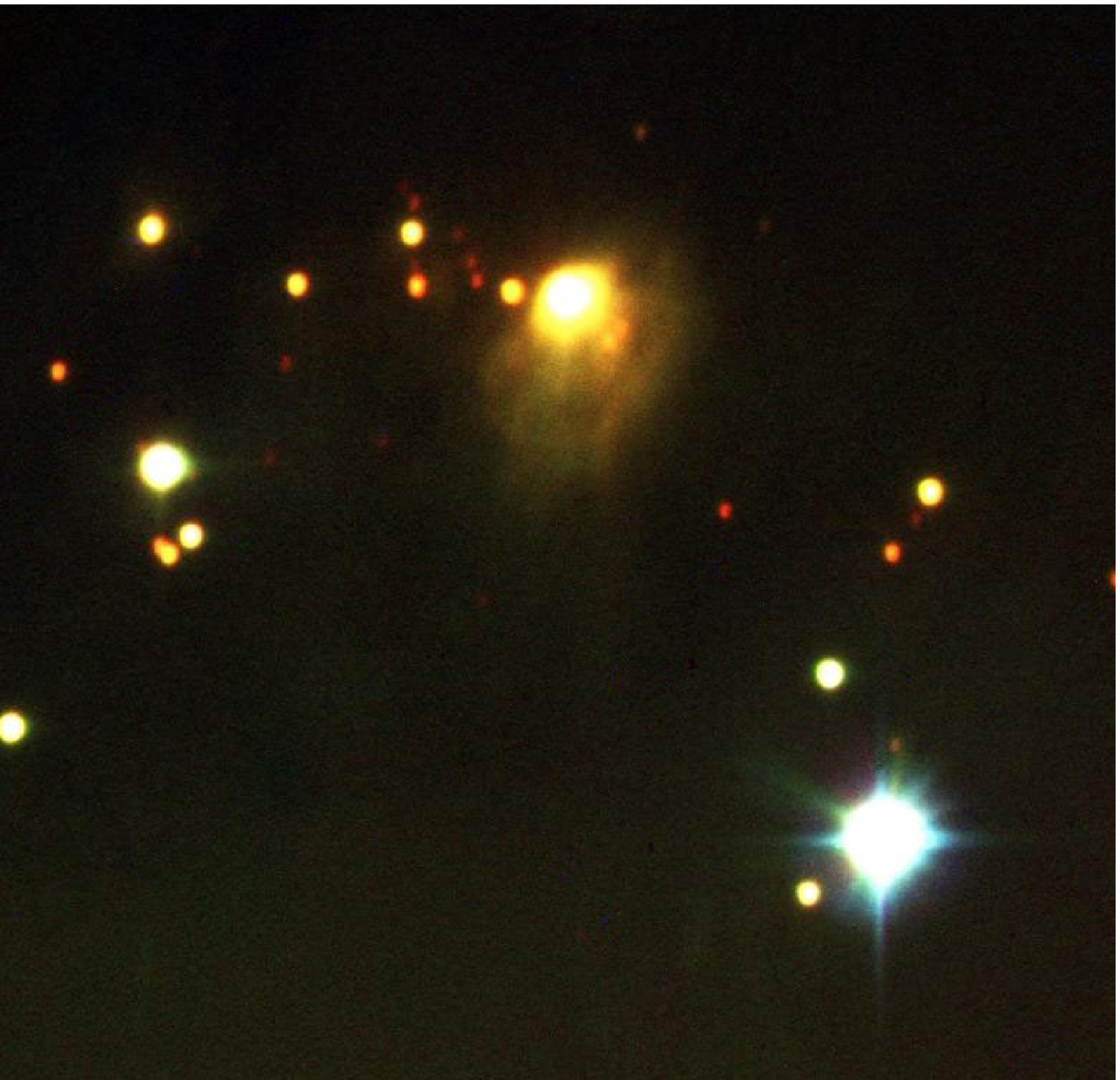}
\caption{Color images of V2493 Cyg obtained with the 2-m RCC telescope of NAO Rozhen, Left: before the outburst on 16 August 2007, 
   Right: after the outburst on 3 August 2013.\label{fig1}}
\end{figure}

The large amplitude outburst of V2493 Cyg was registered in 2010~\cite{Sem10,Mil11}.
In a few months, the brightness of the star increases by more than 4 stellar magnitudes in R-band.
V2493 Cyg is located in a region of active star formation, the dark clouds (so-called ``Gulf of Mexico'') between NGC 7000 and IC 5070. 
\textls[-10]{Simultaneously with the optical outburst appearance of a small reflection nebula surrounding V2493 Cyg was observed (Figure~\ref{fig1}).}
The eruptive star received designation V2493 Cyg according to the General catalog of variable stars, but it has been known in previous studies as HBC 722, LkH 188-G4 and PTF10qpf~\cite{Her88,Coh79,Mil11}.

Follow-up photometric observations by Semkov \& Peneva \cite{Sem11}, K{\'o}sp{\'a}l et al. \cite{Kos11} and Lorenzetti et al. \cite{Lor12} register an ongoing increase in brightness in the optical and near-infrared regions.
Significant changes in the profiles and intensity of the spectral lines are registered in high and low resolution spectroscopic observations \cite{Mun10, Mil11, Lee11, Lee15, Lor12, Sem10, Sem12}.
Miller et al. \cite{Mil11}, K{\'o}sp{\'a}l et al. \cite{Kos11} and Gramajo et al. \cite{Gra14} studied the spectral energy distribution (SED) of the star before and during the eruption. 
The authors found that prior the outburst V2493 Cyg was a Class II young stellar object (YSO) according to the evolutionary sequence proposed by Adams et al. \cite{Ada87}.
The Class II YSOs are most often associated with Classical T Tauri stars, in contrast to the Class I that are younger deeply embedded sources, seen only in the infrared spectral region.
The upper limit of the mass of the circumstellar disc has been studied by Dunham et al. \cite{Dun12} and K{\'o}sp{\'a}l et al. \cite{Kos16}.
The results indicate that the disk surrounding V2493 Cyg has a relatively low mass (0.01-0.02 $M_{\odot}$) for an object of FUor type.
Green et al. \cite{Gre13} have found a 5.8-day and 1.28-day periods in the $r$-band light curve of the star during several observation sessions in 2011-2012. 
The first period is explained by the stellar rotation and the second by Keplerian rotation at the inner radius of the accretion disk.

The unusual photometric behavior of V2493 Cyg is also manifested in subsequent observations~\cite{Bae15,Kos16,Sem14,Sem17}. 
During the period April 2013--October 2016 the star keeps its maximum brightness showing a little bit fluctuations around it.
Unlike other known FUor objects, V2493 Cyg reaches maximum brightness, remains in a photometric plateau without evidences of a decrease in brightness. 
The latest published photometric data on V2493 Cyg confirm the diversity in the light curves of the FUor objects. 
Our knowledge of the processes occurring during the FUor outbursts is still incomplete and it is necessary to collect more data from regular photometric monitoring of such objects. 
 
%%%%%%%%%%%%%%%%%%%%%%%%%%%%%%%%%%%%%%%%%%
\section{Observations}

The present paper is a continuation of our photometric study of V2493 Cyg before and during the outburst~\cite{Sem10,Sem12,Sem14,Sem17}. 
We present recent $BVRI$ data from photometric observations of the star obtained during the period November 2016--February 2021.
The photometric observations of V2493 Cyg were performed with the 2-m Ritchey-Chr\'{e}tien-Coud\'{e} and the 50/70/172 cm Schmidt telescopes of the National Astronomical Observatory Rozhen (Bulgaria).  
Observations were performed with three types of the CCD camera---Vers Array 1300B and ANDOR iKon-L BEX2-DD at the 2-m RCC telescope, and FLI PL16803 at the 50/70 cm Schmidt telescope.
The technical characteristics and optical specifications of the cameras used are listed in Table~\ref{tab1}. 
A standard set of Johnson--Cousins filters were used for all the frames that were obtained. 
Twilight flat fields in each filter were obtained every clear evening. 
All frames obtained with the ANDOR and Vers Array cameras are bias subtracted and flat fielded. 
CCD frames obtained with the FLI PL16803 camera are dark subtracted and flat fielded.

% start a new page without indent 4.6cm

\end{paracol}
\nointerlineskip
\begin{specialtable}[H]
\setlength{\tabcolsep}{3.1mm}
\widetable
\caption{CCD cameras used, technical characteristics and optical specifications.\label{tab1}}
\begin{tabular}{llllllllllll}
  \toprule
%            \noalign{\smallskip}
 \multirow{2}{*}{\textbf{Telescope}\vspace{-4pt}}  & \multirow{2}{*}{\textbf{CCD Camera Type}\vspace{-4pt}}& \textbf{Size} & \textbf{Field}  &\textbf{Pixel Size}&\textbf{Scale} & \textbf{RON}        &\textbf{Gain} \\
 &                &\textbf{(px)}      &\textbf{(arcmin)}&\textbf{(microns)  } &\textbf{(''/px)}&\textbf{(}\boldmath{$e^-$}\textbf{rms)}&\textbf{(}\boldmath{$e^-$}\textbf{/ADU)}\\  
%            \noalign{\smallskip}
      \midrule
%            \noalign{\smallskip}
2~m RCC& Vers Array 1300B&1340~$\times$~1300& 5.8~$\times$~5.6&20.0&0.26&2.00&1.0\\
2~m RCC& ANDOR iKon-L&2048~$\times$~2048& 6.0~$\times$~6.0&13.5&0.17&6.90&1.1\\
Schmidt& FLI PL16803&4096~$\times$~4096& 73.8~$\times$~73.8&9.0&1.08&9.00&1.0\\
%\noalign{\smallskip}
\bottomrule
         \end{tabular}
\end{specialtable}
\begin{paracol}{2}
%\linenumbers
\switchcolumn

All the data were analyzed using the same aperture, which was chosen as 4 arc seconds radius.
The background annulus is taken from 13 arc seconds to 19 arc seconds in order to minimize the light from the surrounding nebula and avoid contamination from nearby stars. 
As references, we used the $BVRI$ comparison sequence of fifteen stars in the field around V2493 Cyg published in Semkov~et~al.~\cite{Sem10}. 
In this way we provided a maximum consistency of the photometric results obtained during the various stages of the photometric observations.
The results of our photometric CCD monitoring of V2493 Cyg are summarized in Table~\ref{tab2}.  
The columns provide the date and Julian date (JD) of observation, $IRVB$ magnitudes of V2493 Cyg, the telescope and CCD camera used. 

\clearpage
\end{paracol}
\nointerlineskip
\begin{specialtable}[H]
\setlength{\tabcolsep}{3.9mm}
\widetable 
\caption{Photometric CCD observations of V2493 Cyg obtained during the period November 2016--February 2021.\label{tab2}}
\begin{tabular}{llllllll}
\toprule
%%\noalign{\smallskip}  
\textbf{Date} \hspace{1.5cm} &	\textbf{JD (24\ldots)} \hspace{2mm}	&	\textbf{I}	\hspace{8mm} & \textbf{R} \hspace{8mm} & \textbf{V} \hspace{8mm} & \textbf{B} \hspace{8mm} & \textbf{Telescope} \hspace{1mm} & \textbf{CCD}	\\
%\noalign{\smallskip}  
\midrule
%\endfirsthead
%\caption{continued.}\\
%\hline\hline
%\noalign{\smallskip}  
%Date \hspace{1.5cm} &	JD (24...) \hspace{2mm}	&	I	\hspace{8mm} & R \hspace{8mm} & V \hspace{8mm} & B \hspace{8mm} & Telescope \hspace{1mm} & CCD	\\
%\noalign{\smallskip}  
%\hline
%\noalign{\smallskip}  
%\endhead
%\hline
%\endfoot
%\noalign{\smallskip}
5 November 2016 %MDPI: we change the date format, please confirm.
  & 57698.240 %MDPI: please confirm whether need to add a comma on 5-digits numbers.
 & 11.30 & 12.49 & 13.54 & 15.09 & Sch & FLI\\
21 November 2016 & 57714.254 & 11.39 & 12.60 & 13.65 & 15.22 & 2~m & VA\\
22 November 2016  & 57715.227 & 11.39 & 12.60 & 13.68 & 15.23 & 2~m & VA\\
23 November 2016  & 57716.245 & 11.40 & 12.63 & 13.72 & 15.25 & 2~m & VA\\
1 January 2017 & 57755.211 & 11.31 & 12.50 & 13.54 & 15.10 & Sch & FLI\\
2 January 2017  & 57756.224 & 11.28 & 12.44 & 13.48 & 15.03 & Sch & FLI\\
28 January 2017  & 57782.197 & 11.34 & 12.44 & 13.65 & 15.13 & 2~m & VA\\
30 January 2017  & 57784.201 & 11.23 & 12.47 & 13.52 & 15.09 & 2~m & VA\\
31 January 2017  & 57785.198 & 11.26 & 12.51 & 13.55 & 15.11 & 2~m & VA\\
1 February 2017  & 57786.199 & 11.27 & 12.46 & 13.54 & 15.06 & 2~m & VA\\
17 February 2017  & 57801.622 & 11.29 & 12.49 & 13.53 & 15.07 & Sch & FLI\\
5 March 2017  & 57817.575 & 11.34 & 12.52 & 13.57 & 15.12 & Sch & FLI\\
2 April 2017  & 57845.561 & 11.26 & 12.45 & 13.50 & 15.06 & Sch & FLI\\
3 April 2017  & 57846.589 & 11.27 & 12.45 & 13.50 & 15.05 & Sch & FLI\\
2 May 2017  & 57875.503 & 11.34 & 12.55 & 13.61 & 15.25 & 2~m & VA\\
18 May 2017  & 57892.397 & 11.37 & 12.57 & 13.61 & 15.16 & Sch & FLI\\
19 May 2017  & 57893.407 & 11.39 & 12.61 & 13.71 & 15.26 & 2~m & VA\\
30 May  2017  & 57904.414 & 11.31 & 12.48 & 13.54 & 15.11 & Sch & FLI\\
12 July 2017 & 57947.471 & 11.24 & 12.42 & 13.49 & 15.05 & Sch & FLI\\
1 August 2017  & 57967.414 & 11.16 & 12.33 & 13.39 & 14.99 & Sch & FLI\\
2 August 2017  & 57968.300 & 11.16 & 12.35 & 13.41 & 15.00 & Sch & FLI\\
3 August 2017  & 57969.298 & 11.19 & 12.37 & 13.45 & 15.05 & Sch & FLI\\
12 August 2017  & 57978.471 & 11.24 & 12.42 & 13.49 & 15.05 & Sch & FLI\\
14 September 2017  & 58011.297 & 11.20 & 12.41 & 13.46 & 15.04 & Sch & FLI\\
15 September 2017  & 58012.315 & 11.19 & 12.36 & 13.44 & 15.01 & Sch & FLI\\
16 September 2017  & 58013.301 & 11.20 & 12.39 & 13.45 & 15.03 & Sch & FLI\\
12 October 2017 & 58039.286 & 11.25 & 12.46 & 13.50 & 15.09 & Sch & FLI\\
14 October 2017  & 58041.283 & 11.26 & 12.45 & 13.50 &       & 2~m & VA\\
16 October 2017  & 58043.267 & 11.27 & 12.49 & 13.55 & 15.13 & Sch & FLI\\
16 October 2017 & 58043.297 & 11.30 & 12.50 & 13.57 & 15.11 & 2~m & VA\\
17 October 2017  & 58044.283 & 11.31 & 12.54 & 13.60 & 15.17 & Sch & FLI\\
18 October 2017  & 58045.380 & 11.28 & 12.53 & 13.59 & 15.15 & Sch & FLI\\
22 October 2017  & 58080.270 & 11.27 & 12.48 & 13.53 & 15.08 & Sch & FLI\\
23 October 2017  & 58081.276 & 11.26 & 12.49 & 13.54 & 15.11 & Sch & FLI\\
21 December 2017  & 58109.298 & 11.21 & 12.40 & 13.46 & 15.03 & Sch & FLI\\
25 December 2014  & 58113.216 & 11.26 & 12.48 & 13.55 & 15.13 & Sch & FLI\\
26 December 2017  & 58114.251 & 11.27 & 12.49 & 13.56 & 15.12 & Sch & FLI\\
9 April 2018  & 58217.594 & 11.24 & 12.43 & 13.49 & 15.07 & Sch & FLI\\
10 April 2018  & 58218.582 & 11.21 & 12.41 & 13.46 & 15.04 & Sch & FLI\\
12 April 2018 & 58220.537 & 11.18 & 12.37 & 13.43 & 14.98 & Sch & FLI\\
8 June 2018 & 58278.415 & 11.17 & 12.34 & 13.41 & 14.96 & Sch & FLI\\
11 June 2018  & 58281.452 & 11.19 & 12.38 & 13.43 & 14.97 & 2~m & ANDOR\\
12 July 2018  & 58312.406 & 11.20 & 12.39 & 13.43 & 14.99 & Sch & FLI\\
16 July 2018  & 58316.367 & 11.24 & 12.44 & 13.51 & 15.08 & Sch & FLI\\
9 August 2018  & 58340.357 & 11.19 & 12.39 & 13.47 & 15.06 & Sch & FLI\\
11 August 2018  & 58342.356 & 11.17 & 12.37 & 13.45 & 15.03 & Sch & FLI\\
12 August 2018  & 58343.369 & 11.14 & 12.32 & 13.39 & 14.97 & Sch & FLI\\
13 August 2018  & 58344.348 & 11.16 & 12.37 & 13.45 & 15.04 & Sch & FLI\\
15 August 2018  & 58346.299 & 11.18 & 12.40 & 13.45 & 15.03 & 2~m & ANDOR\\
1 September 2018  & 58363.323 & 11.22 & 12.41 & 13.49 & 15.08 & Sch & FLI\\
2 September 2018 & 58364.302 & 11.22 & 12.42 & 13.48 & 15.06 & Sch & FLI\\
17 October 2018 & 58409.267 & 11.16 & 12.36 & 13.44 & 15.04 & Sch & FLI\\
\bottomrule
\end{tabular}
\end{specialtable}

\begin{specialtable}[H] \ContinuedFloat
\setlength{\tabcolsep}{3.9mm}
\widetable 
\caption{{\em Cont.}\label{tab2}}
\begin{tabular}{llllllll}
\toprule
%%\noalign{\smallskip}  
\textbf{Date} \hspace{1.5cm} &	\textbf{JD (24\ldots)} \hspace{2mm}	&	\textbf{I}	\hspace{8mm} & \textbf{R} \hspace{8mm} & \textbf{V} \hspace{8mm} & \textbf{B} \hspace{8mm} & \textbf{Telescope} \hspace{1mm} & \textbf{CCD}	\\
%\noalign{\smallskip}  
\midrule
5 November 2018  & 58428.256 & 11.14 & 12.34 & 13.40 & 14.99 & Sch & FLI\\
12 November 2018  & 58435.235 & 11.17 & 12.38 & 13.45 & 15.04 & Sch & FLI\\
8 January 2019 & 58492.201 & 11.34 & 12.57 & 13.66 & 15.29 & Sch & FLI\\
5 March 2019  & 58547.603 & 11.25 & 12.45 & 13.53 & 15.14 & Sch & FLI\\
29 April 2019  & 58603.459 & 11.31 & 12.51 & 13.59 & 15.20 & Sch & FLI\\
1 May 2019  & 58604.511 & 11.28 & 12.48 & 13.56 & 15.17 & Sch & FLI\\
2 May 2019 & 58606.430 & 11.32 & 12.51 & 13.60 & 15.28 & 2~m & ANDOR\\
3 May 2019 & 58606.532 & 11.35 & 12.57 & 13.64 & 15.23 & Sch & FLI\\
30 June 2019  & 58665.394 & 11.28 & 12.48 & 13.55 & 15.15 & Sch & FLI\\
1 July 2019  & 58666.411 & 11.24 & 12.45 & 13.54 & 15.14 & Sch & FLI\\
2 July 2019  & 58667.397 & 11.26 & 12.49 & 13.52 & 15.19 & Sch & FLI\\
25 July 2019 & 58690.305 & 11.30 & 12.50 & 13.56 & 15.24 & 2~m & ANDOR\\
26 July 2019  & 58691.414 & 11.32 & 12.54 & 13.57 & 15.21 & 2~m & ANDOR\\
27 July 2019  & 58692.366 & 11.28 & 12.48 & 13.55 & 15.19 & 2~m & ANDOR\\
8 August 2019  & 58704.349 & 11.28 & 12.50 & 13.59 & 15.21 & Sch & FLI\\
9 August 2019  & 58705.378 & 11.32 & 12.55 & 13.63 & 15.24 & Sch & FLI\\
10 August 2019  & 58706.378 & 11.37 & 12.61 & 13.70 & 15.34 & Sch & FLI\\
11 August 2019  & 58707.453 & 11.34 & 12.55 & 13.64 & 15.29 & Sch & FLI\\
30 August 2019  & 58726.344 & 11.38 & 12.59 & 13.64 & 15.28 & 2~m & ANDOR\\
31 August 2019  & 58727.382 & 11.39 & 12.61 & 13.68 & 15.31 & 2~m & ANDOR\\
1 September 2019  & 58728.423 & 11.40 & 12.62 & 13.74 & 15.33 & 2~m & ANDOR\\
2 September 2019  & 58729.336 & 11.35 & 12.58 & 13.68 & 15.35 & 2~m & ANDOR\\
3 September 2019  & 58729.540 & 11.36 & 12.63 & 13.73 & 15.34 & Sch & FLI\\
3 September 2019  & 58730.340 & 11.36 & 12.61 & 13.73 & 15.42 & 2~m & ANDOR\\
3 September 2019  & 58730.443 & 11.37 & 12.66 & 13.78 & 15.39 & Sch & FLI\\
1 October 2019 & 58758.343 & 11.38 & 12.67 & 13.78 & 15.40 & Sch & FLI\\
2 October 2019  & 58759.331 & 11.39 & 12.66 & 13.77 & 15.40 & Sch & FLI\\
15 January 2020  & 58864.214 & 11.28 & 12.42 & 13.58 & 15.33 & Sch & FLI\\
16 January 2020  & 58865.221 & 11.28 & 12.42 & 13.57 & 15.29 & Sch & FLI\\
18 January 2020 & 58867.184 & 11.29 & 12.41 & 13.55 & 15.29 & 2~m & ANDOR\\
21 January 2020  & 58870.213 & 11.30 & 12.48 & 13.63 & 15.36 & Sch & FLI\\
23 May 2020 & 58993.435 & 11.15 & 12.29 & 13.42 & 15.13 & Sch & FLI\\
9 July 2020  & 59040.356 & 11.17 & 12.31 & 13.45 & 15.16 & Sch & FLI\\
10 July 2020  & 59041.391 & 11.18 & 12.34 & 13.47 & 15.17 & Sch & FLI\\
11 July 2020& 59042.382 & 11.20 & 12.36 & 13.50 & 15.20 & Sch & FLI\\
28 July 2020  & 59059.461 & 11.09 & 12.24 & 13.37 & 15.07 & Sch & FLI\\
29 July 2020  & 59060.369 & 11.13 & 12.28 & 13.40 & 15.12 & Sch & FLI\\
13 August 2020  & 59075.324 & 11.10 & 12.31 & 13.41 & 15.19 & 2~m & ANDOR\\
23 August 2020 & 59085.287 & 11.15 & 12.34 & 13.48 & 15.20 & Sch & FLI\\
8 September 2020 & 59101.417 & 11.11 & 12.27 & 13.41 & 15.13 & Sch & FLI\\
9 September 2020 & 59102.337 & 11.08 & 12.27 & 13.35 & 15.09 & 2~m & ANDOR\\
10 September 2020& 59103.346 & 11.07 & 12.28 & 13.37 & 15.05 & 2~m & ANDOR\\
11 September 2020  & 59104.380 & 11.12 & 12.28 & 13.42 & 15.12 & Sch & FLI\\
12 September 2020& 59105.310 & 11.11 & 12.28 & 13.41 & 15.12 & Sch & FLI\\
15 September 2020 & 59108.334 & 11.12 & 12.27 & 13.41 & 15.10 & Sch & FLI\\
16 September 2020  & 59109.306 & 11.11 & 12.26 & 13.40 & 15.12 & Sch & FLI\\
13 October 2020 & 59136.229 & 11.12 & 12.31 & 13.45 & 15.14 & Sch & FLI\\
23 October 2020  & 59146.319 & 11.14 & 12.30 & 13.44 & 15.15 & Sch & FLI\\
20 November 2020 & 59174.225 & 11.08 & 12.25 & 13.37 & 15.08 & Sch & FLI\\
22 November 2020 & 59176.231 & 11.14 & 12.30 & 13.43 & 15.14 & Sch & FLI\\
23 November 2020  & 59177.216 & 11.13 & 12.29 & 13.43 & 15.12 & Sch & FLI\\
5 January 2021 & 59220.184 & 11.01 & 12.21 & 13.34 & 15.04 & 2~m & ANDOR\\
%4 February 2021  & 59250.196 & 10.96 & 12.14 & 13.25 & 15.06 & 2~m & ANDOR\\
%5 February 2021 & 59251.205 & 10.90 & 12.08 & 13.29 & 15.08 & 2~m & ANDOR\\
%12 February 2021  & 59258.221 & 11.05 & 12.21 & 13.38 &       & Sch & FLI\\
\bottomrule
\end{tabular}
\end{specialtable}

\begin{specialtable}[H] \ContinuedFloat
\setlength{\tabcolsep}{3.9mm}
\widetable 
\caption{{\em Cont.}\label{tab2}}
\begin{tabular}{llllllll}
\toprule
%%\noalign{\smallskip}  
\textbf{Date} \hspace{1.5cm} &	\textbf{JD (24\ldots)} \hspace{2mm}	&	\textbf{I}	\hspace{8mm} & \textbf{R} \hspace{8mm} & \textbf{V} \hspace{8mm} & \textbf{B} \hspace{8mm} & \textbf{Telescope} \hspace{1mm} & \textbf{CCD}	\\
%\noalign{\smallskip}  
\midrule
4 February 2021  & 59250.196 & 10.96 & 12.14 & 13.25 & 15.06 & 2~m & ANDOR\\
5 February 2021 & 59251.205 & 10.90 & 12.08 & 13.29 & 15.08 & 2~m & ANDOR\\
12 February 2021  & 59258.221 & 11.05 & 12.21 & 13.38 &       & Sch & FLI\\
\bottomrule
\end{tabular}
\end{specialtable}

\begin{figure}[H]	
\widefigure
\includegraphics[width=17 cm]{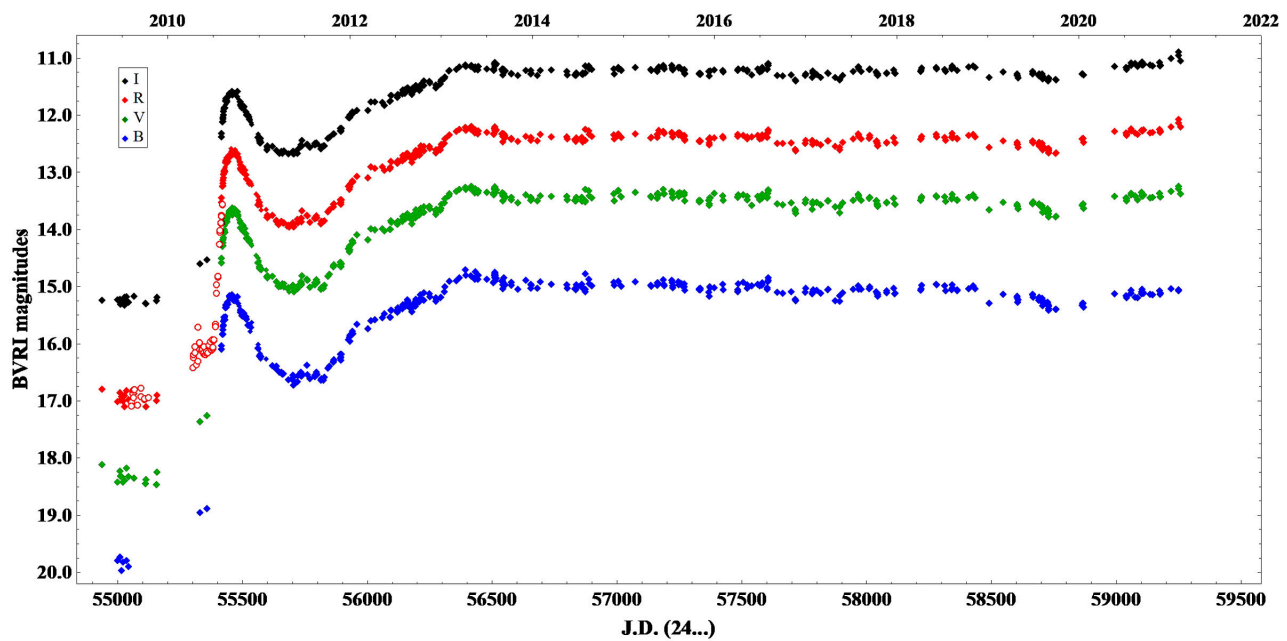}
\caption{$BVRI$ light curves of V2493 Cyg for the period from June 2008 till February 2021.} 
\label{fig2}
\end{figure}

%\end{paracol}
\begin{paracol}{2}
%\linenumbers
\switchcolumn

\vspace{-6pt}
The typical instrumental errors in the reported magnitudes are $0.01$ for $I$ and $R$-band data, 0.01--0.02 for $V$-band, and 0.01--0.03 for $B$-band~\cite{Sem12}.
The typical seeing for the Rozhen observatory is about 2 arc seconds, which is almost always less than the aperture used.
Usually, at least two images are obtained every night in each filter and the table shows the average of them.
In very few cases, there is a difference in the measured values of the two images, greater than 0.01 mag., which shows the good accuracy of our results. 

%%%%%%%%%%%%%%%%%%%%%%%%%%%%%%%%%%%%%%%%%%
\section{Results}

Data collected from photometric observations of V2493 Cyg show that the outburst registered in 2010 has been going on for at least eleven years.
The $BVRI$ light-curves of V2493 Cyg during the period June 2008--February 2021 are plotted in Figure~\ref{fig2}.
The filled diamonds represent our CCD observations (Semkov~et~al.~\cite{Sem10,Sem12,Sem14,Sem17} and the present paper),
and the open circles observations from the 48 inch Samuel Oschin telescope at Palomar Observatory~\cite{Mil11}. 
% start a new page without indent 4.6cm
\vspace{-6pt}
\end{paracol}
\nointerlineskip
\begin{paracol}{2}
%\linenumbers
\switchcolumn

Our $BVRI$ photometric data obtained before the outburst displayed variations with amplitudes less than 1 stellar magnitude in all pass-bands~\cite{Sem12}.
Such variability is characteristic of weak line T Tauri stars and the most common reason is the rotation of the star with asymmetric distribution of cool spots~\cite{Her94,Her07}.
Data collected by Miller~et~al.~\cite{Mil11} indicate that the outburst began sometime before May 2010, and reached the first maximum in brightness during September--October 2010. 
Since October 2010, a slow fading was observed and up to May 2011 the star brightness decreased by 1.4 magnitudes in $V$-band. 
Such a double maximum is not typical for the other FUors.
Occasionally, short dips in brightness are observed after the maximum phase, but they are usually caused by the star being obscured by clouds of dust.
In the case of V2493 Cyg, the most probable reason for the double maximum is variability in the accretion rate.

During the period from May 2011 till October 2011 no significant changes in the brightness of the star were observed, its brightness remains at 3.3 magnitudes in $V$-band, above the quiescence level. 
But since the autumn of 2011, another light increase has begun and the star became brighter by 1.8 magnitudes in $V$-band until the spring of 2013.
During the period April 2013--February 2021 the star remains at its maximum brightness showing a little bit fluctuations around it.

Simultaneously with the increase in brightness of V2493 Cyg its color changes significantly, the star becomes considerably bluer. 
While the both color indexes $V$-$I$ and $R$-$I$ decreased, the $B$-$V$ index remained relatively constant before and during the outburst (Figure~\ref{fig3}).
After six years without changing the values of the color indexes, from 2018 both indexes $V$-$I$ and $R$-$I$ show a gradual increase, the star begins to turn red again. 
This is an evidence of exiting from the maximum phase of the FUor outburst.
But we can't rule out new increases in brightness in the future.
\vspace{-6pt}
\begin{figure}[H]
\includegraphics[width=13 cm]{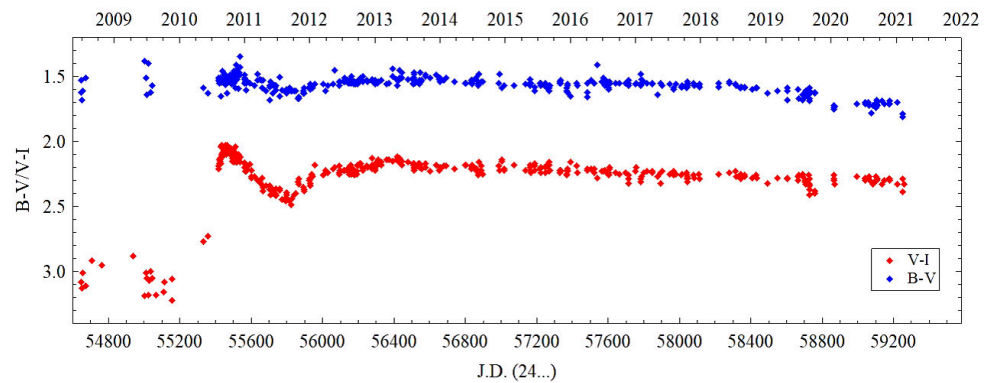}
\caption{$BVRI$ Color evolution of V2493 Cyg from June 2008 till February 2021.\label{fig3}}
\end{figure} 

On the other hand, there is a significant difference in the values of $V$-$I$ and $R$-$I$ indexes during the two peaks of brightness (Figure~\ref{fig4}). 
During the first local maximum (August 2010--May 2011) the star was bluer than during the second maximum (June 2011--February 2021). 
The difference in the $V$-$I$ indexes at the time of the two peaks in brightness reaches 0.2 magnitudes for the same values of $V$ magnitude.
Such a phenomenon can be explained by a gradual expansion of the emitting region around the star.

\vspace{-6pt}
\begin{figure}[H]
\includegraphics[width=10 cm]{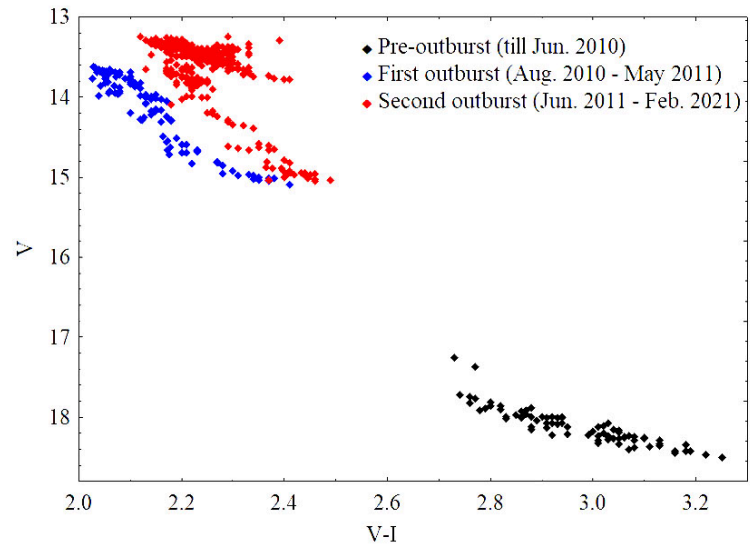}
\caption{$V/V$-$I$ diagram from our $V$ and $I$ photometric data.\label{fig4}}
\end{figure} 

We carried out a periodicity search in the light variations of V2493 Cyg by the software package Period04~\cite{Len05}. 
Such studies are usually severely hampered by the large amplitude photometric variability of the FUors due to the variable accretion rate.
In the case of V2493 Cyg, we have been helped by the long photometric plateau during the recent years.
Therefore, we considered the star’s data received after February 2013, when the brightness of V2493 Cyg varies around some intermediate level in maximal light. 
Our time-series analysis indicates a 914~$\pm$~10 day period. 
The periodogram of the star and its phase-folded $V$ light curve is displayed in Figure~\ref{fig5}.

\begin{figure}[H]
\includegraphics[width=6.5 cm]{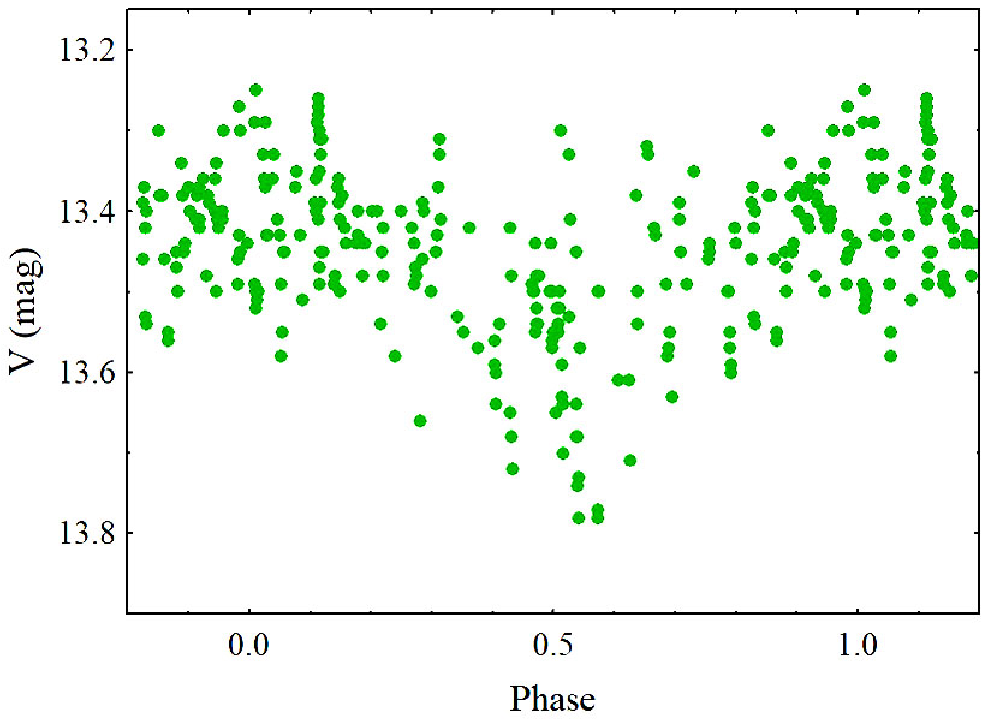}
\includegraphics[width=6.5 cm]{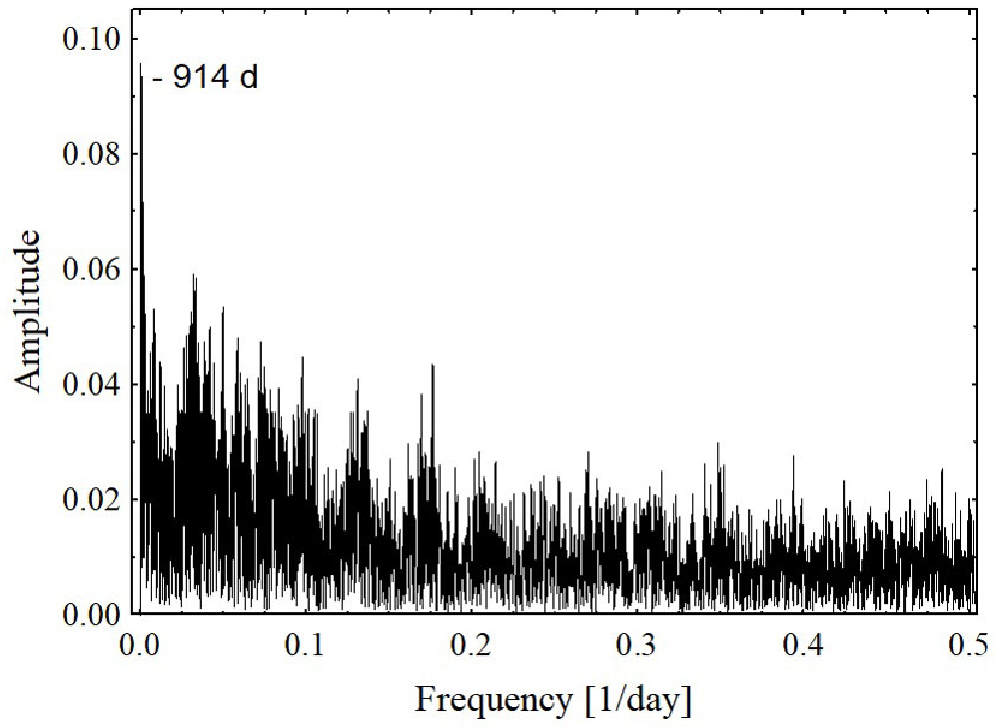}
\caption{The phase-folded $V$ light curve and periodogram of V2493 Cyg for the period February 2013--February 2021.\label{fig5}}
\end{figure}

\section{Discussion}

V2493 Cyg is one of the few FUor objects for which we have spectral and photometric observations before the outburst. 
The spectrum of the star before the outburst shows very intense emission lines of hydrogen, and this classifies it as a T Tauri star~\cite{Coh79,Sem10}.
After the outburst, the spectrum of the star is dominated by absorption lines with strong P Cyg profiles of H-alpha and Na I D lines~\cite{Mun10, Mil11, Lee11, Lor12, Sem12}.
The deep and high velocity absorption lines are interpreted as evidence of a strong outflow driven by the central object. 

In addition to the 5.8-day and 1.28-day periods discovered by Green~et~al.~\cite{Gre13}, we also found a periodicity of 914~$\pm$~10 days.  
A similar periodicity of several hundred days was found in the study of the photometric plateau of the FUor object V1057 Cyg~\cite{Cla05,Sab21}.
In the case of V2493 Cyg we consider that such a periodicity can be explained by dust structures remaining from star formation in orbit around the star.
Our arguments for this assumption are that the minimum is relatively narrow and not very deep. 
A similar effect would be caused by diffuse clouds of dust, which do not have a separate central concentration.
The reason for this periodicity cannot be proved by a change in color indices due to the small amplitude.
By analogy with other FUors such as V1057 Cyg, it justifies such an assumption.

Over the last thirty years, we have undertaken intensive optical photometric monitoring of some unexplored FUor-type objects \cite{Sem21a}.
Since these objects are located in the regions of star formation, we have the opportunity to study the photometric behavior of a large number of PMS stars.
Our data can also be used to detect new FUor or EXor eruptions, to study the light curve and to classify the observed outburst. 
FUors and EXors are the most distinguishable young variable objects due to the large amplitudes and eruptions that last for years. 
Due to the small number of known FUor objects, we do not have sufficient statistics on their photometric evolution during the outburst.

Attempts to classify objects of FUor type on the basis of their photometric properties have so far been unsuccessful. 
The comparison of the light curves of the known FUor objects shows that they are very different from each other and very rarely recur.
Even the first discovered, so-called "classical" FUors (FU Ori, V1057 Cyg and V1515 Cyg) show very different rates of increase and decrease in brightness \cite{Cla05}.
The variety of light curves increases even more with the number of well-studied FUor objects. 
FUors light curves are usually asymmetric, with rapid increases and gradual decreases in brightness. 
Some objects show a very rapid increase in brightness over several months or a year, such as FU Ori, V1057 Cyg and V2493 Cyg \cite{Cla05,Kop13,Sem12}. 
In other cases, such as V1515 Cyg, V1735 Cyg, V733 Cep and V900 Mon, the increase in brightness can last for several years and even reach 20-30 years \cite{Cla05,Pen09,Pen10,Sem21b}. 

A similar variety is observed at the time of decline in brightness.
Usually the decline in brightness takes several decades and will probably take up to a century.
But there are objects in which a relatively rapid decrease in brightness is observed.  
For example, V960 Mon, in which the brightness decreases by 2 stellar magnitudes in the $ V $-band for a period of about five years \cite{Tak20}. 
In the case of V582 Aur two deep decreases in brightens by about 3 magnitudes in $R$-band have been observed separated by a five-year period \cite{Sem13,Abr18}.
In the case of V582 Aur, two deep declines in brightens of about 3 stellar magnitudes in the $R$-band were observed, five years apart. 
But there are also objects that for a period of several decades practically do not change their brightness, as in the case of V1735 Cyg \cite{Pen09,Sem21a} or Parsamian 21 \cite{Sem21a,Sem18}.
In this respect, the light curve of the FUor object V733 Cep is unique, which is approximately symmetrical. 
The rate of increase and the rate of decrease in brightness are almost identical \cite{Pen10}. 
This diversity of photometric properties strongly supports the suggestion that FUor objects are not a homogeneous group and that the causes of this phenomenon may be several different mechanisms \cite{Vor21}. 

\section{Conclusions}

We support the idea that the type of light curve of FUor objects can be evidence of a specific cause of their outburst. 
Trying to compare the available photometric data for V2493 Cyg with the other FUors, we register the following similarities. 
The fast rise in brightness for a period of two years is similar to that of the objects FU Ori, V1057 Cyg and V582 Aur.
At the same time for more than seven years, the brightness of the object remains constant at the level of the maximum like V1735 Cyg and Parsamian 21.  
For now, there are no indications of a decrease in brightness like other FUor objects. 

The results of our studies show that V2493 Cyg is one of the most interesting FUor objects.
V2493 Cyg is the best studied spectroscopically and photometrically FUor object before and after the outburst.
During the outburst, the star was observed in a wide range from the far infrared region to X-rays.
Our results show the need for systematical photometric monitoring of FUor objects.
Continuation of spectral and photometric observations of V2493 Cyg can give us even more important information about the processes of outbursts of FUor type. 

\vspace{6pt}
%%%%%%%%%%%%%%%%%%%%%%%%%%%%%%%%%%%%%%%%%%
\authorcontributions{ %MDPI: For research articles with several authors, a short paragraph specifying their individual contributions must be provided. The following statements should be used ``Conceptualization, X.X. and Y.Y.; methodology, X.X.; software, X.X.; validation, X.X., Y.Y. and Z.Z.; formal analysis, X.X.; investigation, X.X.; resources, X.X.; data curation, X.X.; writing---original draft preparation, X.X.; writing---review and editing, X.X.; visualization, X.X.; supervision, X.X.; project administration, X.X.; funding acquisition, Y.Y. All authors have read and agreed to the published version of the manuscript.'', please turn to the  \href{http://img.mdpi.org/data/contributor-role-instruction.pdf}{CRediT taxonomy} for the term explanation. Authorship must be limited to those who have contributed substantially to the work~reported.
}

\funding{This work was partly supported by the Bulgarian Scientific Research Fund of the Ministry of Education and Science under the grants DN 18-13/2017 and DN 18-10/2017 and by funds of the project RD-08-125/2021 of the University of Shumen.}

\institutionalreview{Not applicable.}

\informedconsent{Not applicable.}

\dataavailability{The data underlying this article are available in the article.} 

\acknowledgments{This research has made use of the NASA's Astrophysics Data System Abstract Service, the SIMBAD database and the VizieR catalogue access tool, operated at CDS, Strasbourg,~France.}

\conflictsofinterest{The authors declare no conflict of interest.}

%%%%%%%%%%%%%%%%%%%%%%%%%%%%%%%%%%%%%%%%%%
%% Only for journal Encyclopedia
%\entrylink{The Link to this entry published on the encyclopedia platform.}

%%%%%%%%%%%%%%%%%%%%%%%%%%%%%%%%%%%%%%%%%%
%% Optional
\abbreviations{Abbreviations}{The following abbreviations are used in this manuscript:\\

\noindent 
\begin{tabular}{@{}ll}
YSO & Young stellar object \\
PMS & Pre-main sequence\\
FUor & Object from FU Orionis type\\
EXor & Object from EX Lupi type\\
$M_{\odot}$ & Solar mass\\ 
SED & Spectral energy distribution\\
NAO & National Astronomical Observatory\\
JD &  Julian date\\
\end{tabular}}

%%%%%%%%%%%%%%%%%%%%%%%%%%%%%%%%%%%%%%%%%%
\end{paracol}
\reftitle{References}


\begin{thebibliography}{999}
% Reference 1

\bibitem[Herbig (1989)]{Her89}
Herbig, G.H. FU Orionis eruptions. In \emph{European Southern Observatory Conference and Workshop Proceedings, Low Mass Star Formation and Pre-Main Sequence Objects}; Reipurth, B., Ed.; European Southern Observatory, Garching, Germany, 1989; pp. 233--246

\bibitem[Herbig (1989)]{Her77}
Herbig, G.H. Eruptive phenomena in early stellar evolution. {\em Astrophys. J.} {\bf 1977}, {\em 217}, 693--715.

\bibitem[Hartmann (1996)]{Har96}
Hartmann, L.; Kenyon, S.J. The FU Orionis Phenomenon. {\em Annu. Rev. Astron. Astrophys.} {\bf 1996}, {\em 34}, 207--240.

\bibitem[Reipurth \& Aspin (2010)]{Rei10}
Reipurth, B.; Aspin, C. FUors and Early Stellar Evolution. In \emph{Evolution of Cosmic Objects through Their Physical,
Proceedings of the Conference Dedicated to Viktor Ambartsumian's 100th Anniversary, Byurakan, Armenia, 15--18 September 2008};  Harutyunian, H.A.,  Mickaelian, A.M., Terzian, Y., Eds.; Publishing House of NAS RA, Yerevan, Armenia, 2010; pp. 19--38

\bibitem[Wachmann (1954)]{Wac54}
Wachmann, A.A. Das bisherige Verhalten von FU Orionis. {\em Z. Astrophys.} {\bf 1954}, {\em 35}, 74--89.

\bibitem[Ambartsumian (1971)]{Amb71}
Ambartsumian, V.A. Fuors. {\em Astrophysics} {\bf 1971}, {\em 7}, 331--339.

\bibitem[Audard (2014)]{Aud14}
Audard, M.; {\'A}brah{\'a}m, P.; Dunham, M.M.; Green, J.D.; Grosso, N.; Hamaguchi, K.; Kastner, J.H.; K{\'o}sp{\'a}l, {\'A}.; Lodato, G.; Romanova, M.M.; et~al. Episodic Accretion in Young Stars, In {\em Protostars and Planets VI}; Beuther, H., Klessen, R.S., Dullemond, C.P., Henning, T., Eds.; University of Arizona Press: Tucson, AZ, USA, {2014}; pp. 387--410.

\bibitem[Connelley (2018)]{Con18}
Connelley, M.S.; Reipurth, B. A Near-infrared Spectroscopic Survey of FU Orionis Objects. {\em Astrophys. J.} {\bf 2018}, {\em 861}, 145.

\bibitem[Zhu (2009)]{Zhu09}
Zhu, Z.; Hartmann, L.; Gammie, C.; McKinney, J.C. Two-dimensional Simulations of FU Orionis Disk Outbursts. {\em Astrophys. J.} {\bf 2009}, {\em 701}, 620--634.

\bibitem[Lodato (2004)]{Lod04}
Lodato, G.; Clarke, C.J. Massive planets in FU Orionis discs: implications for thermal instability models. {\em Mon. Not. R. Astron. Soc.} {\bf 2004}, {\em 353}, 841--852.

\bibitem[Reipurth (2004)]{Rei04}
Reipurth, B.; Aspin, C. The FU Orionis Binary System and the Formation of Close Binaries. {\em Astrophys. J. Lett.} {\bf 2004}, {\em 608}, L65--L68.

\bibitem[Pfalzner (2008)]{Pfa08}
Pfalzner, S. Encounter-driven accretion in young stellar clusters---A connection to FUors? {\em Astron. Astrophys.} {\bf 2008}, {\em 492}, 735--741.

\bibitem[Vorobyov (2005)]{Vor05}
Vorobyov, E.I.; Basu, S. The Origin of Episodic Accretion Bursts in the Early Stages of Star Formation. {\em Astrophys. J. Lett.} {\bf 2005}, {\em 633}, L137--L140.

\bibitem[Vorobyov (2021)]{Vor21}
Vorobyov, E.I.; Elbakyan, V.G.; Liu, H.B.; Takami, M. Distinguishing between different mechanisms of FU-Orionis-type luminosity outbursts. {\em Astron. Astrophys.} {\bf 2021}, {\em 647}, A44.

\bibitem[Semkov (2010)]{Sem10}
Semkov, E.H.; Peneva, S.P.; Munari, U.; Milani, A.; Valisa, P. The large amplitude outburst of the young star HBC 722 in NGC 7000/IC 5070, a new FU Orionis candidate. {\em Astron. Astrophys.} {\bf 2010}, {\em 523}, L3 .

\bibitem[Miller (2011)]{Mil11} 
Miller, A.A.; Hillenbrand, L.A.; Covey, K.R. Poznanski, D.; Silverman, J.M.; Kleiser, I.K.W.; Rojas-Ayala, B.; Muirhead, P.S.; Cenko, S.B.; Bloom, J.S. Evidence for an FU Orionis-like Outburst from a Classical T Tauri Star. {\em Astrophys. J.} {\bf 2011}, {\em 730}, 80.

\bibitem[Herbig (1988)]{Her88} 
Herbig, G.H.;  Bell, K.R.; Robbin, K. Third Catalog of Emission-Line Stars of the Orion Population. {\em Lick Obs. Bull.} {\bf 1988}, 1111.
\bibitem[Cohen (1979)]{Coh79} 
Cohen, M.; Kuhi, L.V. Observational studies of pre-main-sequence evolution. {\em Astrophys. J. Suppl. Ser.} {\bf 1979}, {\em 41}, 743--843.

\bibitem[Semkov (2011)]{Sem11}      
Semkov, E.; Peneva, S. The new FUor star HBC 722---One year after the outburst. {\em Bulg. Astron. J.} {\bf 2011}, {\em 17}, 88--95.

\bibitem[K{\'o}sp{\'a}l (2011)]{Kos11}
K{\'o}sp{\'a}l, {\'A}.; {\'A}brah{\'a}m, P.; Acosta-Pulido, J.A.;  Arévalo Morales, M.J.; Carnerero, M.I.; Elek, E.; Kelemen, J.; Kun, M.; Pál, A.; Szakáts, R.; et~al. The outburst and nature of two young eruptive stars in the North America/Pelican Nebula Complex. {\em Astron. Astrophys.} {\bf 2011}, {\em 527}, A133.

\bibitem[Lorenzetti (2012)]{Lor12} 
Lorenzetti, D.; Antoniucci, S.; Giannini, T.; Li Causi, G.; Ventura, P.; Arkharov, A.A.; Kopatskaya, E.N.; Larionov, V.M.; Di Paola, A.; Nisini, B. On the Nature of EXor Accretion Events: An Infrequent Manifestation of a Common Phenomenology? {\em Astrophys. J.} {\bf 2012}, {\em 749}, 188.

\bibitem[Munari (2010)]{Mun10} 
Munari, U.; Milani, A.; Valisa P.; Semkov, E. Spectroscopic confirmation of HBC 722 as a new FU Orionis star in NGC 7000. {\em  Astron. Telegr.} {\bf 2010}, {\em 2808}. 

\bibitem[Lee (2011)]{Lee11}
Lee, J.-E.; Kang, W.; Lee, S.-G.; Sung, H.-I.; Lee, B.-C.; Sung, H.S.; Green, J.D.; Jeon, Y.-B. High Resolution Optical Spectra of HBC 722 after Outburst. {\em J. Korean Astron. Soc.} {\bf 2011}, {\em 44}, 67--72.

\bibitem[Lee (2015)]{Lee15} 
Lee, J.-E.; Park, S.; Green, J.D.; Cochran, W.D.; Kang, W.; Lee, S.-G.; Sung, H.-I. High Resolution Optical and NIR Spectra of HBC 722. {\em Astrophys. J.} {\bf 2015}, {\em 807}, 84.

\bibitem[Semkov (2012)]{Sem12}  
Semkov, E.H.; Peneva, S.P.; Munari, U.; Tsvetkov, M.K.; Jurdana-Šepić, R.; de Miguel, E.; Schwartz, R.D.; Dimitrov, D.P.; Kjurkchieva, D.P.; Radeva, V.S. Optical photometric and spectral study of the new FU Orionis object V2493 Cygni (HBC 722). {\em Astron. Astrophys.} {\bf 2012}, {\em 542}, A43.

\bibitem[Gramajo (2014)]{Gra14}
Gramajo, L.V.; Rod{\'o}n, J.A.; G{\'o}mez, M. Spectral Energy Distribution Analysis of Class I and Class II FU Orionis Stars. {\em Astron. J.} {\bf 2014}, {\em 147}, 140. 

\bibitem[Adams (1987)]{Ada87} 
Adams, F.C.; Lada, C.J.; Shu, F.H. Spectral Evolution of Young Stellar Objects. {\em Astrophys. J.} {\bf 1987}, {\em 312}, 788--806.

\bibitem[Dunham (2012)]{Dun12}
Dunham, M.M.; Arce, H.G.; Bourke, T.L.; Chen, X.; van Kempen, T.A.; Green, J.D. Revealing the Millimeter Environment of the New FU Orionis Candidate HBC722 with the Submillimeter Array. {\em Astrophys. J.} {\bf 2012}, {\em 755}, 157.

\bibitem[K{\'o}sp{\'a}l (2016)]{Kos16}
K{\'o}sp{\'a}l, {\'A}.; {\'A}brah{\'a}m, P.; Acosta-Pulido, J.A.; Dunham, M.M.; García-Álvarez, D.; Hogerheijde, M.R.; Kun, M.; Moór, A.; Farkas, A.; Hajdu, G.;~et~al. Multiwavelength study of the low-luminosity outbursting young star HBC 722. {\em Astron. Astrophys.} {\bf 2016}, {\em 596}, A52.

\bibitem[Green (2013]{Gre13}
Green, J.D.; Robertson, P.; Baek, G.; Pooley, D.; Pak, S.; Im, M.; Lee, J.-E.; Jeon, Y.; Choi, C.; Meschiari, S. Variability at the Edge: Optical Near/IR Rapid-cadence Monitoring of Newly Outbursting FU Orionis Object HBC 722. {\em Astrophys. J.} {\bf 2013}, {\em 764}, 22.

\bibitem[Baek (2015]{Bae15} 
Baek, G.; Pak, S.; Green, J.D.; Meschiari, S.; Lee, J.-E.; Meschiari, S.; Lee, J.-E.; Jeon, Y.; Choi, C.; Im, M.; et~al. Color Variability of HBC 722 in the Post-Outburst Phases. {\em Astron. J.} {\bf 2015}, {\em 149}, 73 .

\bibitem[Semkov (2014)]{Sem14}
Semkov, E.H.; Peneva, S.P.; Ibryamov, S.I.; Dimitrov, D.P. The unusual photometric behavior of the new FUor star V2493 Cyg (HBC 722). {\em Bulg. Astron. J.} {\bf 2014}, {\em 20}, 59--67.

\bibitem[Semkov (2017)]{Sem17}
Semkov, E.H.; Peneva, S.P.; Ibryamov, S.I. Photometric and spectroscopic study of the new FUor star V2493 Cyg. {\em Bulg. Astron. J.} {\bf 2017}, {\em 26}, 57--66.

\bibitem[Herbst (1994)]{Her94}
Herbst, W.; Herbst, D.K.; Grossman, E.J.; Weinstein, D. Catalogue of UBVRI Photometry of T Tauri Stars and Analysis of the Causes of Their Variability. {\em Astron. J.} {\bf 1994}, {\em 108}, 1906--1923.

\bibitem[Herbst (2007)]{Her07} 
Herbst, W.; Eisl\"{o}ffel, J.; Mundt, R.; Scholz, A. The Rotation of Young Low-Mass Stars and Brown Dwarfs, In {\em Protostars and Planets V}; Reipurth, B., Jewitt, D., Keil, K., Eds.; University of Arizona Press: Tucson, AZ, USA, {2017}; pp. 297--311 .

\bibitem[Lenz (2005)]{Len05}
Lenz, P.; Breger, M. Period04 User Guide. {\em Commun. Asteroseismol.} {\bf 2005}, {\em 146}, 53--136.

\bibitem[Clarke (2005)]{Cla05}
Clarke, C.; Lodato, G.; Melnikov, S.Y.; Ibrahimov, M.A. The photometric evolution of FU Orionis objects: disc instability and wind-envelope interaction. {\em Mon. Not. R. Astron. Soc.} {\bf 2005}, {\em 361}, 942--954.

\bibitem[Szabó (2021)]{Sab21}
Szabó, Z.M.; Kóspál, Á.; Ábrahám, P.; Park, S.; Siwak, M.; Green, J.D.; Moór, A.; Pál, A.; Acosta-Pulido, J.A.; Lee, J.-E.;~et~al. A Study of the Photometric and Spectroscopic Variations of the Prototypical FU Orionis-type Star V1057 Cyg. {\em Astrophys. J.} {\bf 2021},  {\em 917}, 80.

\bibitem[Semkov (2021)]{Sem21a}
Semkov, E.; Peneva, S.; Ibryamov, S.; Munari, U.; Mito, H. Optical Light Curves of the FUor and FUor-like Objects. In Proceedings of the Contributions of the Conference: Star Formation:  From Clouds to Discs, a Tribute to the Career of Lee Hartmann, Dublin, Ireland, 18--21 October 2021. Available online: \url{https://zenodo.org/record/5577408} (accessed on October 19, 2021).


\bibitem[Kopatskaya (2013)]{Kop13}
Kopatskaya, E.N.; Kolotilov, E.A.; Arkharov, A.A. Photometric behaviour of the FU Orionis type star, V1057 Cygni, during the last 25 years. {\em Mon. Not. R. Astron. Soc.} {\bf 2013}, {\em 434}, 38--45.

\bibitem[Peneva (2009)]{Pen09}
Peneva, S.P.; Semkov, E.H.; Stavrev, K.Y. Photometric study of the FUor star V 1735 Cyg (Elias 1--12). {\em Astrophys. Space Sci.} {\bf 2009}, {\em 323}, 329--335.

\bibitem[Peneva (2010)]{Pen10}
Peneva, S.P.; Semkov, E.H.; Munari, U.; Birkle, K. A long-term photometric study of the FU Orionis star V 733 Cephei. {\em Astron. Astrophys.} {\bf 2010}, {\em 515}, A24.

\bibitem[Semkov (2021)]{Sem21b}
Semkov, E.H.; Peneva, S.P.; Ibryamov, S.I. Long-term Optical Photometric Monitoring of the FUor Star V900 Mon. {\em Serb. Astron. J.} {\bf 2021}, {\em 202}, 31--38.

\bibitem[Takagi (2020)]{Tak20}
Takagi, Y.; Honda, S.; Arai, A.; Takahashi, J.; Oasa, Y.; Itoh, Y. Revealing the Spectroscopic Variations of FU Orionis Object V960 Mon with High-resolution Spectroscopy.
{\em Astrophys. J.} {\bf 2020}, {\em 904}, 53 .

\bibitem[Semkov (2013)]{Sem13}  
Semkov, E.H.; Peneva, S.P.; Munari, U.; Dennefeld, M.; Mito, H.; Dimitrov, D.P.; Ibryamov, S.; Stoyanov, K.A. Photometric and spectroscopic variability of the FUor star V582 Aurigae. {\em Astron. Astrophys.} {\bf 2013}, {\em 556}, A60.

\bibitem[{\'A}brah{\'a}m (2018)]{Abr18}
{\'A}brah{\'a}m, P.; K{\'o}sp{\'a}l, {\'A}.; Kun, M.; Fehér, O.; Zsidi, G.; Acosta-Pulido, J.A.; Carnerero, M.I.; García-Álvarez, D.; Moór, A.; Cseh, B.;~et~al. An UXor among FUors: Extinction-related Brightness Variations of the Young Eruptive Star V582 Aur. {\em Astrophys. J.} {\bf 2018}, {\em 853}, 28.

\bibitem[Semkov (2018)]{Sem18}
Semkov, E.; Ibryamov, S.; Peneva, S.; Mutafov, A. Long-term photometric monitoring of FUor and FUor-like objects. {\em Commun. Byurakan Astrophys. Obs.} {\bf 2018}, {\em 65}, 240--248.


\end{thebibliography}
\end{document}